\documentclass[11pt]{article}
\usepackage{cite}
\usepackage{amsmath,amsfonts,amssymb}%,calrsfs}
\usepackage[small,bf,hang]{caption}
\usepackage{slashed}
\usepackage{mathabx}

\def\hybrid{
        \topmargin -20pt
        \oddsidemargin 0pt
        \headheight 0pt \headsep 0pt
        \textwidth 6.25in % A4 paper
        \textheight 9.5in % A4 paper
        \marginparwidth .875in
        \parskip 5pt plus 1pt \jot = 1.5ex}

\hybrid

 \csname
@addtoreset\endcsname{equation}{section}

\def\moth{\mathsurround=0pt}
\newdimen\zo \zo=0pt

\def\tick{\leaders\hrule height 0.5ex depth 0pt \hskip 0.5pt}
\def\upboxfill{$\moth \setbox\zo\hbox{\tick}%
  \hskip 3pt\hbox to 0pt{$\tick$\hss}\hrulefill \hbox to 7.5pt{$\tick$\hss}$}

\def\dtick{\leaders\hrule height .34pt depth 0.5ex \hskip 0.5pt}
\def\downboxfill{$\moth \setbox\zo\hbox{\dtick}%
  \hskip 2pt\hbox to 0pt{$\dtick$\hss}\hrulefill \hbox to 2pt{$\dtick$\hss}$}

\def\bec{\begin{center}}
\def\ec{\end{center}}

\def\be{\begin{equation}}
\def\ee{\end{equation}}
\def\bea{\begin{eqnarray}}
\def\eea{\end{eqnarray}}
\def\ba{\begin{array}}
\def\ea{\end{array}}

\begin{document}

\begin{titlepage}
\rightline{}
%\rightline{\tt }
\rightline{\tt  LMU-ASC 44/13}
%\rightline\today
\rightline{July 2013}
\begin{center}
\vskip 1.0cm
{\Large \bf {Gauge theory of Kaluza-Klein and winding modes}}\\
\vskip 1.4cm
{\large {Olaf Hohm${}^1$ and Henning Samtleben${}^2$}}
\vskip 1cm
{\it {${}^1$Arnold Sommerfeld Center for Theoretical Physics}}\\
{\it {Theresienstrasse 37}}\\
{\it {D-80333 Munich, Germany}}\\
olaf.hohm@physik.uni-muenchen.de
\vskip 0.2cm
{\it {${}^2$Universit\'e de Lyon, Laboratoire de Physique, UMR 5672, CNRS}}\\
{\it {\'Ecole Normale Sup\'erieure de Lyon}}\\
{\it {46, all\'ee d'Italie, F-69364 Lyon cedex 07, France}}\\
henning.samtleben@ens-lyon.fr

\vskip 1.5cm
{\bf Abstract}
\end{center}

\vskip 0.2cm

\noindent
\begin{narrower}
We perform a Kaluza-Klein inspired rewriting of double field theory 
by splitting the coordinates into `compact' and `non-compact' directions. 
There is \textit{no truncation} of the compact coordinates or their duals, and 
so this formulation is manifestly $O(d,d)$ invariant, with $d$ the number 
of compact directions. The action can serve as starting point for arbitrary Kaluza-Klein 
ans\"atze. For a torus background the theory describes the full tower
of Kaluza-Klein modes or, in the dual frame, of the winding modes. 
The Kaluza-Klein vector is a gauge field for the 
duality-covariantized Courant bracket algebra rather than a Lie algebra.
Gauge covariance requires the inclusion of the 2-form gauge potential 
descending from the Kalb-Ramond field, leading to 
a structure resembling the tensor hierarchy of gauged supergravity.

\end{narrower}

\end{titlepage}

\newpage

 \tableofcontents
 
 \bigskip
 \bigskip
 \bigskip
 
 \section{Introduction}
The celebrated T-duality property of closed string theory implies that toroidal 
backgrounds $T^d$ related via the non-compact group $O(d,d,\mathbb{Z})$
are physically equivalent. Another manifestation of this duality is the emergence 
of the continuous global symmetry group $O(d,d,\mathbb{R})$ 
in the Kaluza-Klein compactification of the corresponding supergravity on a torus. 
More precisely, 
this holds if the theory is truncated to the massless modes only; once the massive 
Kaluza-Klein modes are taken into account, the $O(d,d,\mathbb{R})$ symmetry is broken.
In fact, including massive Kaluza-Klein modes amounts to keeping the  
radius of the torus finite, and in this case string theory requires the inclusion of 
winding modes. In string theory this restores the $O(d,d,\mathbb{Z})$ 
T-duality symmetry, mapping Kaluza-Klein and winding modes into each other, 
but in conventional supergravity T-duality is no longer visible. 
One purpose of the present paper is to exhibit the $O(d,d)$ structure 
in more conventional field theory, using double field theory. 

Double field theory (DFT) has been constructed from various angles in 
\cite{Siegel:1993th,Hull:2009mi,Hull:2009zb,Hohm:2010jy,Hohm:2010pp}. 
(See also the earlier work in  \cite{Tseytlin:1990va,Duff:1989tf,Kugo:1992md,Siegel:1993xq,Hitchin:2004ut,Gualtieri:2003dx,Gualtieri:2007bq} and 
\cite{Hohm:2010xe,Hohm:2011ex,Hohm:2011dz,Hohm:2011zr,Hohm:2011dv,Hohm:2011cp,Hohm:2011nu,
Hohm:2011si,Hohm:2012gk,Hohm:2012mf,Jeon:2010rw,Jeon:2011cn,Jeon:2012kd,Coimbra:2011nw,Jeon:2011sq,
Jeon:2012hp,Aldazabal:2011nj,Geissbuhler:2011mx,Grana:2012rr,Dibitetto:2012rk,Berman:2013uda,Andriot:2012wx,Andriot:2012an,Geissbuhler:2013uka} 
for generalizations and applications. Reviews have appeared in \cite{Hohm:2011gs,Zwiebach:2011rg,Aldazabal:2013sca,Berman:2013eva}.)
It is written in terms of doubled coordinates and assembles  
them into an $O(D,D)$ vector. The most geometric form of the DFT action reads \cite{Hohm:2010pp} 
 \be\label{HDFTaction}
  S_{\rm DFT}  \ = \ \int d^{2D}X\,e^{-2d}\,{\cal R}({\cal H},d)\;, 
 \ee
written in terms of the generalized scalar curvature ${\cal R}$ that is a function of 
the dilaton density $d$ and the familiar generalized metric 
${\cal H}$ encoding metric  and B-field. Truncating the dependence on the dual 
coordinates, the action reduces to the standard space-time action of closed string theory,  
 \bea\label{original}
  S \ = \ \int d^Dx \sqrt{-g}e^{-2\phi}\left[R+4(\partial\phi)^2-\frac{1}{12}H^2\right]\,.
 \eea
This idea of a doubled geometry is well-motivated from string theory
on a torus background, where the new coordinates are dual to winding modes.
Thus, string theory suggests that a doubled geometry should be appropriate for 
backgrounds of the form $\mathbb{R}^{n-1,n}\times T^d$, and the  
construction of the cubic theory from string field theory in \cite{Hull:2009mi} 
employed such a background. Later work in \cite{Hohm:2010jy} 
gave a manifestly background-independent formulation such as used in (\ref{HDFTaction}). 
This formulation features a global 
 continuous $O(D,D)$ symmetry, where $D$ is the total number of space-time dimensions 
 that need not be decomposed into $n$ non-compact and $d$ compact directions.
 Formally, this theory doubles all space-time coordinates, but it is subject to the 
 `strong constraint' which states 
  \be\label{STRONG}
  \eta^{\hat{M}\hat{N}}\partial_{\hat{M}}\partial_{\hat{N}} \ \equiv \ \partial^{\hat{M}}\partial_{\hat{M}} \ = \ 0\;, \qquad 
   \eta^{\hat{M}\hat{N}} \ = \ \begin{pmatrix}    0 & {\bf 1} \\[0.5ex]
  {\bf 1} & 0 \end{pmatrix}\;, 
  \ee
acting on arbitrary fields and products of fields. Here, $\partial_{\hat{M}}$ denotes the 
derivatives dual to the doubled coordinates $X^{\hat{M}}$, combining the space-time 
coordinates $x$ and their duals $\tilde{x}$, and $\eta_{\hat{M}\hat{N}}$ is the 
$O(D,D)$ invariant metric. This constraint is a stronger form of the level-matching 
constraint for the massless sector of string theory on a torus, $L_0-\bar{L}_0=0$,
and it implies that locally the theory depends only on half of the coordinates. 
The full string theory on a torus background only requires the weaker constraint 
that $\partial^{\hat{M}}\partial_{\hat{M}}$ annihilates all fields but not necessarily all products, 
which allows for field configurations that depend even locally both on $x$ and $\tilde{x}$. 
It is a matter of current research to what extent and in which sense the strong constraint (\ref{STRONG}) 
can be relaxed \cite{Hohm:2011cp,Geissbuhler:2011mx,Grana:2012rr}. Throughout this paper we assume the strong constraint,
but comment later on possible relaxations. 

Our aim in this paper is to give DFT in a formulation that is appropriate for Kaluza-Klein 
compactifications, in a sense returning to the original idea of a theory  with 
`non-compact' and `compact' directions. We stress, however, that this does not 
require any particular restriction of the `internal manifold' nor any assumption on the topology of the background. 
Rather, we simply gauge fix the local Lorentz symmetry and split the 
coordinates \textit{without any truncation} in the compact coordinates, writing 
 \be\label{COORdinatesplit}
  x^{\hat{m}} \ = \ (x^{\mu}\, ,\; y^m)\;, \qquad \tilde{x}_{\hat{m}} \ = \ (\tilde{x}_{\mu}\,,\; \tilde{y}_m)\;, \qquad
  \mu \ = \ 0,\ldots,n-1\;, \quad m \ = \ 1,\ldots,d\;, 
 \ee 
where $D=n+d$. Somewhat loosely, we will refer to the coordinates as non-compact and compact, 
although this does not entail any assumption on the topology. 
The local Lorentz group of DFT, which consists of two copies of the usual 
Lorentz group, is then gauge fixed according to 
 \be\label{gaugeFIX}
  O(D-1,1)_L\times O(D-1,1)_R \quad \longrightarrow\quad O(n-1,1)\times O(d)_L\times O(d)_R\;. 
 \ee 
In addition, we assume that the fields are independent of
the dual space-time coordinates $\tilde{x}_{\mu}$, i.e., $\tilde{\partial}^{\mu}=0$, but leaving 
the dependence on $Y^M=(\tilde{y}_m,y^m)$, $M=1,\ldots, 2d$,  and thus the rigid $O(d,d)$ symmetry untouched. 
In this sense our approach is an extension of that of de~Wit-Nicolai, who gave a Kaluza-Klein rewriting 
of 11-dimensional supergravity, but without truncation \cite{deWit:1986mz,Nicolai:1986jk}. Similarly, our formulation contains 
the usual two-derivative space-time theory (\ref{original}) of closed string theory, 
but rewritten by use of a Kaluza-Klein-like decomposition of fields and coordinates, 
thereby still depending generally on $y^m$. However, our formulation goes beyond the 
approach of de~Wit-Nicolai (as applied to string theory) in that we also keep the dependence on the dual coordinates 
$\tilde{y}_m$, thereby arriving at an $O(d,d)$ covariant action. 

This action can now serve as the convenient starting point for an arbitrary Kaluza-Klein ansatz. 
For instance, we may specialize to a torus ansatz and expand the fields in Fourier modes  
$\exp(im_ny^n)$ and $\exp (iw^n\tilde{y}_n)$, thus giving the effective action of the full tower of 
Kaluza-Klein or winding modes. Although by the strong constraint (\ref{STRONG}) so far we can 
only describe either  Kaluza-Klein or winding modes (or any combination obtained 
thereof by an $O(d,d)$ rotation), the action is manifestly $O(d,d)$ covariant.  

As another Kaluza-Klein ansatz we may choose Scherk-Schwarz reductions, which have 
already been studied extensively in the literature on DFT \cite{Aldazabal:2011nj,Geissbuhler:2011mx,Grana:2012rr,Dibitetto:2012rk,Berman:2013uda}, see also~\cite{DallAgata:2007sr}.
In particular, it has been found that the duality-covariant language of DFT is 
perfectly suited in order to reproduce gauged supergravity theories in lower dimensions,  
formulated with the embedding tensor technique~\cite{Nicolai:2000sc,deWit:2002vt}. 
As such, it can reproduce in a transparent manner 
those gauged supergravities that have no conventional uplift to higher-dimensional supergravity 
(and that are related to non-geometric fluxes) --- although not in full generality unless the 
strong constraint is relaxed, as we will discuss in the outlook. 
A crucial ingredient of the embedding tensor formalism and the duality-covariant 
form of gauged supergravity is the so-called \textit{tensor hierarchy} that, in addition to the 
usual (non-abelian) one-form gauge fields, needs to introduce forms of higher rank in order to maintain 
gauge covariance~\cite{deWit:2005hv,deWit:2008ta,Bergshoeff:2009ph}. 
One of the main results of this paper is to show that 
the gauge structure of the DFT action given here
is governed by an analogue 
of the tensor hierarchy \textit{prior} to any Kaluza-Klein compactification. 

In order to explain this point in a little more detail, we recall that in DFT the gauge  
transformations parametrized by a gauge parameter $\Lambda^M$ read, say, on a vector 
\be\label{GennnLie0}
  \delta_{\Lambda}V^M \ = \ 
 \big[\Lambda,V\big]_{ D}^M\ \equiv \ \Lambda^N\partial_N V^M +\big(\partial^M\Lambda_N-\partial_N\Lambda^M\big)V^N\;, 
 \ee  
where we introduced the D-bracket, which is the duality-covariantized extension of the 
Dorfman bracket in generalized geometry. These transformations differ from the usual 
infinitesimal diffeomorphisms given by the Lie derivative and thus represent \textit{generalized 
Lie derivatives}. They close according to the C-bracket, 
$[\delta_{\Lambda_1},\delta_{\Lambda_2}]=-\delta_{[\Lambda_1,\Lambda_2]_{C}}$, 
which is the antisymmetrization of the D-bracket and the duality-covariantized version of 
the Courant bracket in generalized geometry.  Remarkably, this bracket does not define 
a Lie algebra in that the Jacobi identity does not hold \cite{Gualtieri:2003dx,Hull:2009zb}. Now, in the Kaluza-Klein-type 
rewriting of DFT we have a Kaluza-Klein vector $A_{\mu}{}^{M}$, which acts as a 
gauge field for the $\Lambda^M$ gauge transformations,\footnote{The linearized gauge 
transformations of similar fields have also been investigated in \cite{Cederwall:2013oaa}.}  
 \be
   \delta_{\Lambda}A_{\mu}{}^{M} \ = \ \partial_{\mu}\Lambda^M+\big[\Lambda,A_{\mu}\big]_{D}^{M}\;. 
 \ee
This allows for the definition of covariant derivatives $D_{\mu}$, but since neither 
the C- nor D-bracket define a Lie algebra we cannot employ the usual results of 
Yang-Mills theory. In particular, since the C-bracket violates the Jacobi identity,
the naive field strength 
 \be
  F_{\mu\nu}{}^{M} \ = \ \partial_{\mu}A_{\nu}{}^{M}-\partial_{\nu}A_{\mu}{}^{M}-\big[A_{\mu},A_{\nu}\big]_{C}^{M}\;, 
 \ee
is not a covariant object. The failure of $F$ to transform covariantly is, however, of the 
`exact' form $\partial^M\chi$ for some function $\chi$ and therefore we can define a fully covariant 
field strength 
 \be\label{FHatintro}
  {\cal F}_{\mu\nu}{}^{M} \ \equiv \ F_{\mu\nu}{}^{M}-\partial^{M}B_{\mu\nu}\;, 
 \ee  
by introducing a 2-form gauge potential $B_{\mu\nu}$  with an appropriate gauge transformation.  
This is precisely analogous to the tensor hierarchy, where the gauge group structure does not 
define a proper Lie algebra and accordingly higher forms need to be introduced in order to 
guarantee gauge covariance. Luckily, DFT provides precisely such a 2-form, originating from 
the external components of the $B$-field. Similarly, we will have to introduce a 
field strength for this 2-form,  
 \be\label{Gfieldstrength0}
  {\cal H}_{\mu\nu\rho} \ = \ 3\Big(D_{[\mu}B_{\nu\rho]} + A_{[\mu}{}^{N}\partial_{\nu}A_{\rho]N}
  -\frac{1}{3}A_{[\mu N}\big[A_{\nu},A_{\rho]}\big]^N_{ C} \Big)\;, 
 \ee
where a Chern-Simons-type modification is necessary for gauge covariance. 
The appearance of a Chern-Simons 3-form in the curvature of the 2-form is a 
well-known phenomenon in string theory, but here a generalization of 
the Chern-Simons term is needed, based on the (covariantized) Courant bracket. 

With these novel gauge structures we are ready to write the DFT action, with $\tilde{\partial}^{\mu}=0$, 
for the Kaluza-Klein fields 
 \be
\{\;\;  g_{\mu\nu}\;, \;\; B_{\mu\nu}\;, \;\;  \phi\;, \;\;  {\cal H}_{MN}\;, \;\;  A_{\mu}{}^{M}\;\;\} \;, 
 \ee
where ${\cal H}_{MN}$ denotes the $O(d,d)$ matrix encoding the internal components of 
the metric and the $B$-field and $\phi$ is a Kaluza-Klein redefinition of the DFT dilaton. 
The action is 
manifestly invariant under $\Lambda$ transformations and reads 
 \be\label{finalActionIntro}
  \begin{split}
   S \ = \ \int d^nx\, d^{2d}y\,e\,e^{-2\phi}\Big(  &\widehat{R}
   +4g^{\mu\nu}D_{\mu}\phi D_{\nu}\phi -\frac{1}{12}{\cal H}^{\mu\nu\rho}{\cal H}_{\mu\nu\rho}\\
   &+\frac{1}{8}g^{\mu\nu}D_{\mu}{\cal H}^{MN}D_{\nu}{\cal H}_{MN}
   -\frac{1}{4}{\cal H}_{MN}{\cal F}^{\mu\nu M}{\cal F}_{\mu\nu}{}^{N}-V\Big)\;,  
  \end{split}
 \ee  
cf.~(\ref{finalAction}) below. The potential is given by 
 \be
  V(\phi,{\cal H},g) \ = \ -{\cal R}(\phi,{\cal H}) - \frac{1}{4}{\cal H}^{MN}\partial_Mg^{\mu\nu}\,\partial_N g_{\mu\nu} \;, 
 \ee 
with the scalar curvature as in the DFT action (\ref{HDFTaction}), but written for ${\cal H}_{MN}$ and $\phi$.  
This action is a twofold extension of the zero-mode action for the torus compactification of the 
closed string action (\ref{original}), given by Maharana-Schwarz \cite{Maharana:1992my}.
Not only does it preserve the full dependence on all internal coordinates, but also it does so in a duality covariant way,
keeping the full dependence on the doubled internal coordinates $Y^M$.
Correspondingly, all derivatives and curvatures are replaced by covariant derivatives and 
non-abelian field strenghts. Moreover, there is a potential that in Kaluza-Klein language (say, with 
respect to a torus) induces the masses of the higher Kaluza-Klein/winding modes. 

This paper is organized as follows. In sec.~2 we develop systematically the 
tensor hierarchy for the C- and the D-bracket 
by introducing covariant derivatives, curvatures and proving 
Bianchi identities. Then, in sec.~3, we prove starting from DFT and performing the 
Kaluza-Klein-like decomposition that its action can indeed be written in the above form.
We close with a short discussion of the implications for the effective action of 
Kaluza-Klein or winding modes and with a general outlook.

\newpage

\section{The tensor hierarchy for the D- and C-bracket}
In this section we develop the tensor hierarchy based on the gauge structures of 
DFT, which are governed by the D- and C-bracket. We start by recalling some results from DFT and introducing 
the relevant notation.\footnote{Here we write all relations for `un-hatted' $2d$-valued indices $M,N$, in order not to clutter 
the notation; of course, the general DFT relations are also valid for tensors with $2D$ valued indices $\hat{M}$, $\hat{N}$.}
The gauge transformations are governed by generalized Lie derivatives that 
we denote by $\widehat{\cal L}$ and which 
act on a vector according to the D-bracket (\ref{GennnLie0}):
 \be\label{GennnLie}
  \delta_{\Lambda}V^M \ = \ 
  \widehat{\cal L}_{\Lambda}V^M  \ \equiv \ \Lambda^N\partial_N V^M +\big(\partial^M\Lambda_N-\partial_N\Lambda^M\big)V^N\;. 
 \ee  
All indices are raised and lowered with the $O(d,d)$ invariant metric $\eta_{MN}$. 
The action of the generalized Lie derivative on a general tensor is the straightforward 
extension of (\ref{GennnLie}). 
Note that the $O(d,d)$ metric $\eta_{MN}$ itself is invariant under $\widehat{\cal L}_{\Lambda}$. 
Moreover, due to the strong constraint (\ref{STRONG}) a gauge parameter of the form 
$\Lambda^M=\partial^M\chi$ does not generate a gauge transformation and is hence referred to as a 
trivial parameter.  
The generalized Lie derivatives form a closed algebra, 
 \be\label{CBRacket}
  \big[ \widehat{\cal L}_{\Lambda_1}, \widehat{\cal L}_{\Lambda_2}\big] \ = \  \widehat{\cal L}_{[\Lambda_{1},\Lambda_2]_{ C}}\;, 
 \ee
where the bracket is given by the C-bracket 
 \be
  \big[\Lambda_1,\Lambda_2\big]_{ C}^{M} \ \equiv \ \Lambda_1^N\partial_N\Lambda_2^M-\frac{1}{2}\Lambda_{1N}
  \partial^M\Lambda_2^N-(1\leftrightarrow 2)\;.
 \ee
Comparing with (\ref{GennnLie0}) one may check that the C-bracket differs from the D-bracket 
by a total derivative symmetric in the arguments 
 \be\label{SymmC}
  \big[V,W\big]^M_{ C} \ = \   \big[V,W\big]^M_{ D}-\frac{1}{2}\partial^M\big(V^N W_N\big)\;.
 \ee 
We finally note that the Jacobiator of the C-bracket is non-zero, but takes the trivial form \cite{Hull:2009zb}
 \be\label{Jacobiator}
  \big[\big[U,V\big]_{C},W\big]_{C}^{M}+{\rm cycl.} \ = \ \frac{1}{6}\partial^M\left(\big[U,V\big]_{C}^{N}W_{N}+ {\rm cycl.}\right)\;, 
 \ee 
which will be important below when constructing gauge covariant curvatures. 

We now turn to the introduction of gauge fields and invariant field strengths. These will covariantize 
derivatives $\partial_{\mu}$ with respect to the $n$ `non-compact' coordinates, which is necessary 
since the gauge parameters are also functions of $x^{\mu}$, $\Lambda=\Lambda(x,Y)$, and hence the symmetry is 
local with respect to the external space.  
We start with the gauge transformations of  $A_{\mu}{}^{M}$, which in analogy to ordinary 
Yang-Mills theory we define to be 
 \be\label{betterdelta}
  \delta_{\Lambda}A_{\mu}{}^{M} \equiv \ \partial_{\mu}\Lambda^M+\big[\Lambda,A_{\mu}\big]_{D}^{M}
  \ = \ \partial_{\mu}\Lambda^M-\big[A_{\mu},\Lambda\big]_{D}^{M}+\partial^M\big(\Lambda^N A_{\mu N})\;. 
 \ee
Here we used that according to (\ref{SymmC}) the D-bracket is not antisymmetric and so the  
two natural ways of writing the gauge transformations  
as in Yang-Mills theory differ by a total $\partial^M$ derivative. As we will explain below, this difference is 
irrelevant due to an extra shift gauge symmetry on $A_{\mu}{}^{M}$.  
With the gauge field $A_{\mu}{}^{M}$ we can next define a covariant $x^{\mu}$-derivative, 
 \be\label{covDerGen}
  D_{\mu} \ = \ \partial_{\mu}-\widehat{\cal L}_{A_{\mu}}\;.  
 \ee 
Here, the generalized Lie derivative acts in the representation of the object on which $D_{\mu}$
acts.  Thus, for the internal generalized metric ${\cal H}_{MN}$ the covariant derivative is given by  
  \be\label{covM}
    D_{\mu}{\cal H}_{MN} \ = \ \partial_{\mu}{\cal H}_{MN}-A_{\mu}{}^{K}\partial_K {\cal H}_{MN}
    -2\left(\partial_{(M} A_{\mu}{}^{K}-\partial^KA_{\mu (M}\right){\cal H}_{N)K} \;.
   \ee 
Below, we also need the dilaton $e^{-2\phi}$, which is a density with respect to $\Lambda$ 
transformations, and the external vielbein $e_{\mu}{}^{a}$, which is a $\Lambda$-scalar,
 \begin{equation}
  \begin{split}
   \delta_{\Lambda} e^{-2\phi} \ &= \ \Lambda^N\partial_N e^{-2\phi}+\partial_N\Lambda^N e^{-2\phi}\;, \\
   \delta_{\Lambda} e_{\mu}{}^{a} \ &= \ \Lambda^N\partial_N e_{\mu}{}^{a}\;. 
  \end{split}
 \end{equation}   
Thus, their covariant derivatives read 
 \be
 \begin{split}
  D_{\mu}e^{-2\phi} \ &= \ \partial_{\mu}e^{-2\phi}-A_{\mu}{}^{N}\partial_N e^{-2\phi}-\partial_NA_{\mu}{}^{N} e^{-2\phi}\;,\\
  D_{\mu}e_{\nu}{}^{a} \ &= \ \partial_{\mu}e_{\nu}{}^{a}-A_{\mu}{}^{N}\partial_N e_{\nu}{}^{a}\;.
 \end{split}
 \ee

Despite the slightly non-standard form of the gauge transformations of the gauge fields, 
these derivatives are fully covariant under local $\Lambda^M$ transformations. 
To see this consider a generic $O(d,d)$ tensor $W$. We then compute 
 \be
  \begin{split}
   \delta_{\Lambda} (D_{\mu} W) \ &= \ \partial_{\mu}(\widehat{\cal L}_{\Lambda}W)-
   \widehat{\cal L}_{\partial_{\mu}\Lambda+[\Lambda, A_{\mu}]_{C}}W-\widehat{\cal L}_{A_{\mu}}\widehat{\cal L}_{\Lambda}W \\
   \ &= \ \widehat{\cal L}_{\Lambda}(\partial_{\mu}W)+\widehat{\cal L}_{\partial_{\mu}\Lambda}W
    -\widehat{\cal L}_{\partial_{\mu}\Lambda}W-\widehat{\cal L}_{[\Lambda, A_{\mu}]_{C}}W
   -\widehat{\cal L}_{A_{\mu}}\widehat{\cal L}_{\Lambda}W  \\
    \ &= \ \widehat{\cal L}_{\Lambda}(\partial_{\mu}W)-\big[\widehat{\cal L}_{\Lambda},\widehat{\cal L}_{A_{\mu}}\big]W
     -\widehat{\cal L}_{A_{\mu}}\widehat{\cal L}_{\Lambda}W \\
     \ &= \  \widehat{\cal L}_{\Lambda}(\partial_{\mu}W-\widehat{\cal L}_{A_{\mu}}W)
     \ = \ \widehat{\cal L}_{\Lambda}(D_{\mu}W)\;. 
  \end{split}
 \ee  
Here we have written in the first line $\delta A_{\mu}{}^M$ as in (\ref{betterdelta}), but with 
the C-bracket instead of the D-bracket, using that by (\ref{SymmC}) the difference is a trivial 
parameter that does not generate a generalized Lie derivative.   In the second line we used the 
linearity of the generalized Lie derivative in its argument, and in the third line we used (\ref{CBRacket}). 

Next let us attempt to construct a covariant field strength of $A_{\mu}{}^{M}$. In analogy with Yang-Mills theory we set   
 \be\label{fieldstr}
  F_{\mu\nu}{}^{M} \ = \ \partial_{\mu}A_{\nu}{}^{M}-\partial_{\nu}A_{\mu}{}^{M}-\big[A_{\mu},A_{\nu}\big]_{C}^{M}\;. 
 \ee
As usual, the field strength can also be defined through the commutator of covariant derivatives,  
 \be\label{commF}
  \big[D_{\mu},D_{\nu}\big] \ = \ -\widehat{\cal L}_{F_{\mu\nu}}\;, 
 \ee 
as can be easily verified.  
Since the C- or D-bracket do not define a Lie algebra, however, $F_{\mu\nu}{}^{M}$ does not transform fully covariantly. 
An explicit computation shows 
 \be\label{DeltaF}
   \delta_{\Lambda}F_{\mu\nu}{}^{M} \ = \ \widehat{\cal L}_{\Lambda}F_{\mu\nu}{}^{M}+\partial^{M}\big(\partial_{[\mu}\Lambda^N  
     A_{\nu]N}\big)\;.
 \ee
Thus, while $F_{\mu\nu}{}^{M}$ is 
not fully gauge covariant, by the strong constraint it is gauge invariant in terms contracted as $F_{\mu\nu}{}^{M}\partial_{M}$. 

In a next step we introduce a 2-form potential in order to construct a fully covariant field strength. We set  
 \be\label{FHat}
  {\cal F}_{\mu\nu}{}^{M} \ = \ F_{\mu\nu}{}^{M}-\partial^{M}B_{\mu\nu}\;. 
 \ee
From (\ref{DeltaF}) we infer that  
 this modified field strength 
 transforms covariantly, $\delta_{\Lambda}{\cal F}_{\mu\nu}{}^{M}=\widehat{\cal L}_{\Lambda}{\cal F}_{\mu\nu}{}^{M}$,  
if the 2-form transforms as 
 \be\label{BTransF}
  \delta_{\Lambda}B_{\mu\nu} \ = \ \Lambda^N\partial_NB_{\mu\nu}+\partial_{[\mu}\Lambda^{N} A_{\nu]N}\;.
 \ee
The general variation of $B$ can actually be written in a more convenient form as follows. 
First consider the general variation of $F_{\mu\nu}{}^M$ induced by an arbitrary $\delta A_{\mu}{}^{M}$, 
  \be
  \delta F_{\mu\nu}{}^{M} \ = \ \partial_{\mu}(\delta A_{\nu}{}^{M})-\big[ A_{\mu},\delta A_{\nu}\big]_{C}^{M}-(\mu\leftrightarrow \nu)\;. 
 \ee
In conventional Yang-Mills theory we would rewrite this in terms of the covariant derivatives of $\delta A_{\mu}{}^{M}$.
However, since the covariant derivative (\ref{covDerGen}) on a $\Lambda$-vector involves the D-bracket rather than the antisymmetric 
C-bracket, we cannot rewrite this immediately in terms of covariant derivatives. Rather, 
we find with (\ref{SymmC}) 
 \be
  \delta F_{\mu\nu}{}^{M} \ = \ D_{\mu}(\delta A_{\nu}{}^{M})-D_{\nu}(\delta A_{\mu}{}^{M})
  +\partial^M(A_{[\mu}{}^{N}\delta A_{\nu]N})\;. 
 \ee      
If we now introduce a `covariant variation' of the 2-form,  
 \be\label{DeltaB}
  \Delta B_{\mu\nu} \ \equiv \ \delta B_{\mu\nu}-A_{[\mu}{}^{N}\delta A_{\nu]N}\;, 
 \ee
 the general variation of the covariantized field strength (\ref{FHat}) 
 can be written as 
   \be\label{delfullF}
  \delta {\cal F}_{\mu\nu} \ = \ 2D_{[\mu}\, \delta A_{\nu]}{}^{M}-\partial^M\Delta B_{\mu\nu}\;, 
 \ee
which is in precise analogy to the tensor hierarchy~\cite{deWit:2005hv,deWit:2008ta}.  
In terms of this covariant variation of $B_{\mu\nu}$ we can
rewrite the $\Lambda^M$ gauge variation (\ref{BTransF}) of $B_{\mu\nu}$ in a more compact way. 
After a short computation one finds 
 \be\label{DELtaB}
  \Delta B_{\mu\nu} \ = \ -\Lambda^N {\cal F}_{\mu\nu N} +2D_{[\mu}\big(\Lambda^N A_{\nu]N}\big)\;, 
 \ee
where the covariant derivative acts on $\Lambda^N A_N$ as for a $\Lambda$-scalar.   
Next, we can also define a gauge transformation of 
$B_{\mu\nu}$ under a (one-form) parameter $\Lambda_{\mu}$, 
  \be\label{oneformgauge}
  \Delta B_{\mu\nu}  \ = \ 2D_{[\mu}\Lambda_{\nu]}\;, \qquad
  \delta A_{\mu}{}^{M} \  = \ \partial^M\Lambda_{\mu}\;. 
 \ee
The field strength (\ref{FHat}) is invariant under this transformation, as follows from (\ref{delfullF}), 
 \be 
 \delta{\cal F}_{\mu\nu}{}^M \ = \ 2D_{[\mu}\partial^M \Lambda_{\nu]}-2\partial^M D_{[\mu}\Lambda_{\nu]} \ = \ 0\;, 
 \ee
using that $D_{\mu}$ and $\partial^M$ commute when acting on an $O(d,d)$ scalar such as $\Lambda_{\mu}$, 
which can be easily verified.  We note from the second transformation in (\ref{oneformgauge}) that the 
$\Lambda^M$ gauge transformation of $A_{\mu}{}^{M}$ is only determined up to total $\partial^M$ derivatives thanks to the extra $\Lambda_{\mu}$ shift symmetry, cf.\ the discussion after (\ref{betterdelta}). In fact, if we add a field-dependent 
gauge transformation with parameter $\Lambda_{\mu}=-\Lambda^NA_{\mu N}$, we infer from (\ref{betterdelta}) and 
(\ref{DELtaB}) that the $\Lambda^M$ gauge transformations can also be written as 
 \be
  \Delta B_{\mu\nu} \ = \ -\Lambda^N{\cal F}_{\mu\nu N}\;, \qquad 
  \delta A_{\mu}{}^{M} \ = \ D_{\mu}\Lambda^M\;.
 \ee  
 
Another consequence of the extra shift-like symmetry on $A_{\mu}{}^{M}$  is 
the appearance of a `gauge symmetry of gauge symmetries'. To see this, recall that 
the generalized Lie derivative and thus the D-bracket 
is trivial for $\Lambda^M=\partial^M\chi$ and so from (\ref{betterdelta}) 
 \be
  \delta_{\partial\chi}A_{\mu}{}^{M} \ = \ \partial_{\mu}\partial^M\chi
  +\partial^M\Lambda_{\mu}\;.
 \ee 
Thus, no gauge symmetry is generated if we choose the parameters in total as 
 \be\label{trivparamters}
    \Lambda^M \ = \ \partial^M\chi\;, \qquad \Lambda_{\mu} \ = \ -\partial_{\mu}\chi\;.
 \ee     
Similarly, for $\Lambda^M=\partial^M\chi$ we have for the 2-form from (\ref{BTransF})
 \be\label{firstDeltaB}
  \delta_{\partial\chi}B_{\mu\nu} \ = \ \partial_{[\mu}\partial^N\chi\, A_{\nu]N} \;, 
 \ee 
while for $\Lambda_{\mu} =  -\partial_{\mu}\chi$ we have from  (\ref{oneformgauge}) 
 \be
  \delta B_{\mu\nu} \ = \ 2D_{[\mu}\Lambda_{\nu]}+A_{[\mu}{}^{N}\partial_N\Lambda_{\nu]}
  \ = \  2\partial_{[\mu}\Lambda_{\nu]}-A_{[\mu}{}^{N}\partial_N\Lambda_{\nu]}  \ = \ 
  A_{[\mu}{}^{N}\partial_{N}\partial_{\nu]}\chi\;,
 \ee 
which precisely cancels (\ref{firstDeltaB}). Thus, the parameters (\ref{trivparamters}) also do not generate a 
gauge transformation on $B$.   
Again, this has a direct analogue in the tensor hierarchy of gauged supergravity, see eq.~(3.30) in \cite{Samtleben:2008pe}.

Let us now turn to the closure of the gauge algebra, returning to the original form 
of the gauge transformations (\ref{betterdelta}) and 
(\ref{DELtaB}). Unlike in conventional Yang-Mills theory the $\Lambda^M$ transformations do not 
close by themselves, but rather require the extra $\Lambda_{\mu}$ shift symmetry.  
We compute with (\ref{betterdelta}) 
 \be
 \begin{split}
  \big[\delta_{\Lambda_1},\delta_{\Lambda_2}\big]A_{\mu}{}^{M} \ &= \ \delta_{\Lambda_1}\big(\partial_{\mu}\Lambda_2^M+\widehat{\cal L}_{\Lambda_2}A_{\mu}{}^{M}\big)
  -(1\leftrightarrow 2) \\
  \ &= \ \widehat{\cal L}_{\Lambda_2}\big(\partial_{\mu}\Lambda_1^M\big)+\widehat{\cal L}_{\Lambda_2}\widehat{\cal L}_{\Lambda_1}A_{\mu}{}^{M}    
  -(1\leftrightarrow 2) \\
  \ &= \ \big[\Lambda_2,\partial_{\mu}\Lambda_1\big]_C^M+\frac{1}{2}\partial^M\big(\Lambda_2^N\partial_{\mu}\Lambda_{1N}\big)
  +\widehat{\cal L}_{\Lambda_2}\widehat{\cal L}_{\Lambda_1}A_{\mu}{}^{M}    
  -(1\leftrightarrow 2)\\
  \ &= \  \partial_{\mu}\big[\Lambda_2,\Lambda_1\big]_{C}^M+\big[\widehat{\cal L}_{\Lambda_2},\widehat{\cal L}_{\Lambda_1}\big]A_{\mu}{}^{M} 
  +\frac{1}{2}\partial^M\big(\Lambda_2^N\partial_{\mu}\Lambda_{1N}-(1\leftrightarrow 2) \big)\;. 
 \end{split}
 \ee 
Here we rewrote the generalized Lie derivative and thus the D-bracket in terms of the antisymmetric C-bracket, 
using (\ref{SymmC}). Employing now the algebra (\ref{CBRacket}) of generalized Lie derivatives to simplify 
the last line, we infer   
  \be
  \big[ \delta_{\Lambda_1},\delta_{\Lambda_2}\big]A_{\mu}{}^{M} 
  \ = \ \delta_{[\Lambda_2,\Lambda_1]_C}A_{\mu}{}^{M}+\partial^M\Lambda_{12\mu}\;, 
  \qquad \Lambda_{12\mu} \ = \ \frac{1}{2}\Lambda_{2N}\partial_{\mu}\Lambda_1^N -(1\leftrightarrow 2)\;, 
 \ee 
thereby establishing closure. Similarly, one may verify closure on $B_{\mu\nu}$ according to 
the same parameters.

\medskip
Next, we introduce a 3-form field strength for the 2-form. 
The easiest way to derive the field strength is to note that the usual Bianchi identity 
does not hold for $F_{\mu\nu}{}^{M}$. 
Indeed, we compute
 \be
 \begin{split}
  D_{[\mu}F_{\nu\rho]}{}^{M} \ &= \ 
 \partial_{[\mu}F_{\nu\rho]}{}^{M}-\big[A_{[\mu},F_{\nu\rho]}\big]^M_{D} \\
  \ &= \ -2\big[\partial_{[\mu}A_{\nu},A_{\rho]}\big]_{C}^M-\big[A_{[\mu},F_{\nu\rho]}\big]^M_{C}
  -\frac{1}{2}\partial^M\big(A_{[\mu}{}^{N}F_{\nu\rho] N}\big) \\
  \ &= \ \big[A_{[\mu},\big[A_{\nu},A_{\rho]}\big]_C\big]^M_{C}
  -\frac{1}{2}\partial^M\left(A_{[\mu}{}^{N}F_{\nu\rho] N}\right)  \;. 
 \end{split}
 \ee 
Here we inserted the explicit form of the field strength (\ref{fieldstr}), and used in the second line 
the relation between the C- and D-bracket. We can next use that the totally antisymmetrized 
double commutator in the last line equals the Jacobiator (\ref{Jacobiator})  of the C-bracket. 
We finally obtain 
  \be
  D_{[\mu}F_{\nu\rho]}{}^{M} \ = \ -\partial^M \Big(A_{[\mu}{}^{N}\partial_{\nu}A_{\rho]N}
  -\frac{1}{3}A_{[\mu N}\big[A_{\nu},A_{\rho]}\big]^N_{ C}\Big)\;.
 \ee
The failure of $F$ to satisfy the Bianchi identity is again of  `trivial' form, so that when 
contracted with $\partial_M$ it vanishes.  Alternatively, 
the terms on the right-hand side can be absorbed into the field strength of the 2-form. 
We find for the covariantized field strength (\ref{FHat}) 
 \be
  D_{[\mu}{\cal F}_{\nu\rho]}{}^{M} \ = \ D_{[\mu}F_{\nu\rho]}{}^{M} -\partial^M\big(D_{[\mu}B_{\nu\rho]}\big)\;, 
 \ee 
where 
 \be
  D_{\mu}B_{\nu\rho} \ = \ \partial_{\mu}B_{\nu\rho}-A_{\mu}{}^{N}\partial_NB_{\nu\rho}\;, 
 \ee
and we used again that in this case $\partial^M$ and $D_{\mu}$ commute.    
Defining the field strength of the 2-form as 
 \be\label{Gfieldstrength}
  {\cal H}_{\mu\nu\rho} \ = \ 3\Big(D_{[\mu}B_{\nu\rho]} + A_{[\mu}{}^{N}\partial_{\nu}A_{\rho]N}
  -\frac{1}{3}A_{[\mu N}\big[A_{\nu},A_{\rho]}\big]^N_{ C} \Big)\;, 
 \ee
we obtain the modified Bianchi identity 
 \be\label{modBianchi}
   3 D_{[\mu}{\cal F}_{\nu\rho]}{}^{M} +\partial^M {\cal H}_{\mu\nu\rho} \ = \ 0\;.
 \ee  
An explicit computation shows that ${\cal H}_{\mu\nu\rho}$ is fully covariant 
under $\Lambda^M$ transformations
and invariant under $\Lambda_{\mu}$ transformations, i.e., 
 \be
  \delta_{\Lambda} {\cal H}_{\mu\nu\rho} \ = \ \Lambda^M\partial_M {\cal H}_{\mu\nu\rho}\;.
 \ee 
The 3-form field strength also satisfies a generalized Bianchi identity,  
\bea
D_{[\mu}\, {\cal H}_{\nu\rho\sigma]} -\frac34\,{\cal F}_{[\mu\nu}{}^M\,{\cal F}_{\rho\sigma]}{}_M
&=&
0\;,  
\eea 
which follows as an integrability 
condition from (\ref{modBianchi}), using (\ref{commF}). 
 We note that the fact that ${\cal H}_{\mu\nu\rho}$ transforms fully covariantly
  is somewhat different from the tensor hierarchy in gauged supergravity, 
where usually one also has to introduce a 3-form gauge field in the field strength of the 2-form. 
This may suggest that the strong constraint is stronger
than necessary. We will come back to this point in the conclusions.

\section{Kaluza-Klein form of closed string theory action} 

\subsubsection*{The DFT action}
In this section we derive the action (\ref{finalActionIntro}) given in the introduction 
from the DFT action (\ref{HDFTaction}) and 
also show that the gauge transformations emerge from DFT in precisely the way required 
by the tensor hierarchy discussed above. 

We begin with a brief review of DFT.
As we perform a Kaluza-Klein-type rewriting of the action it is convenient to 
employ a frame or vielbein formalism. The original work by Siegel gives such a frame formalism \cite{Siegel:1993th}, 
and it has been shown in \cite{Hohm:2010xe} how to relate it to the generalized metric formulation 
of DFT. In this formalism one works with a frame field $E_{\hat{A}}{}^{\hat{M}}$ (where as above hatted indices 
refer to $2D$-valued indices). The frame field is subject to a tangent space gauge group, which can be as large 
as $GL(D)\times GL(D)$, but for our present purposes it is sufficient to work with two copies of the local 
Lorentz group, which then will be gauge fixed as in (\ref{gaugeFIX}). In the frame formalism of \cite{Siegel:1993th,Hohm:2010xe} 
the two local Lorentz groups can be neatly disentangled by means of a unbarred/barred index notation, 
which is very convenient when introducing fermions \cite{Hohm:2011nu}, 
but then the frame field itself is an $O(D,D)$ group element only up to a similarity transformation. 
Since in this paper we are not interested in fermions it is convenient to 
slightly adapt the frame formalism by working with a frame field that is a proper $O(D,D)$ group element, 
as has been used by Geissb\"uhler \cite{Geissbuhler:2011mx}. 
Thus, in this formulation we take the vielbein to satisfy 
 \be\label{vielbeinconstr}
   E_{\hat{A}}{}^{\hat{M}}E_{\hat{B}}{}^{\hat{N}}\eta_{\hat{M}\hat{N}} \ = \ \eta_{\hat{A}\hat{B}} \;, 
 \ee
i.e., the $O(D,D)$ metric with `flattened indices' takes the same form as with curved indices.    
The vielbein transforms under gauge transformations as 
 \be\label{fuilGAUGE}
  \delta_{\xi}  E_{\hat{A}}{}^{\hat{M}} \ = \ \xi^{\hat{N}}\partial_{\hat{N}}E_{\hat{A}}{}^{\hat{M}}
  +\big(\partial^{\hat{M}}\xi_{\hat{N}}-\partial_{\hat{N}}\xi^{\hat{M}}\big)E_{\hat{A}}{}^{\hat{N}}
  +\lambda_{\hat{A}}{}^{\hat{B}}E_{\hat{B}}{}^{\hat{M}}\;, 
 \ee
where $\lambda$ is the gauge parameter for local $O(D-1,1)_L\times O(D-1,1)_{R}$ transformations. 
The generalized metric can then be defined as 
 \be\label{genHdef}
  {\cal H}_{\hat{M}\hat{N}} \ = \ E_{\hat{M}}{}^{\hat{A}}\,E_{\hat{N}}{}^{\hat{B}}\,\delta_{\hat{A}\hat{B}}\;, 
 \ee
where we refer to eq.~(\ref{deltaDef}) below for the precise meaning of the $\delta$ symbol. 
The generalized metric is invariant under local Lorentz transformations. 
Finally, the theory requires the dilaton density $d$, which transforms as 
 \be\label{dilvar}
  \delta_{\xi} e^{-2d} \ = \  \partial_{\hat{N}}\big(\xi^{\hat{N}}e^{-2d}\big)\;, 
 \ee 
and which is needed to define an invariant integration. 

In order to write the DFT action in this frame formalism we introduce (generalized) 
coefficients of anholonomy \cite{Siegel:1993th,Hohm:2010xe}, 
 \be\label{anholonomy1}
  \widehat{\Omega}_{\hat{A}\hat{B}\hat{C}} \ = \ 3f_{[\hat{A}\hat{B}\hat{C}]}\;, \qquad
  f_{\hat{A}\hat{B}\hat{C}} \ \equiv \ E_{\hat{A}}{}^{\hat{M}}\partial_{\hat{M}}E_{\hat{B}}{}^{\hat{N}}\,E_{\hat{N}\hat{C}} \ = \ 
  -f_{\hat{A}\hat{C}\hat{B}}\;, 
 \ee  
and 
 \be\label{anholonomy2}
  \widehat{\Omega}_{\hat{A}} \ = \ -e^{2 d}\partial_{\hat{M}}\big(E_{\hat{A}}{}^{\hat{M}}e^{-2d}\big)\;.
 \ee 
The DFT action is then given by
 \be\label{DFToncemore}
  S_{\rm DFT} \ = \ \int d^{2D}X \,{\cal L}_{\rm DFT}\;, 
 \ee
where the Lagrangian density reads, up to total derivatives,  \cite{Geissbuhler:2011mx}
 \be\label{DFTLagrangian}
  {\cal L}_{\rm DFT} \ = \ e^{-2d}\Big(\;\frac{1}{4}\delta^{\hat{A}\hat{B}}\,\widehat{\Omega}_{\hat{A}}{}^{\hat{C}\hat{D}}  \,
  \widehat{\Omega}_{\hat{B}\hat{C}\hat{D}}   -\frac{1}{12}\delta^{\hat{A}\hat{B}}\delta^{\hat{C}\hat{D}}\delta^{\hat{E}\hat{F}}\,\widehat{\Omega}_{\hat{A}\hat{C}\hat{E}}  \,
  \widehat{\Omega}_{\hat{B}\hat{D}\hat{F}}   +\delta^{\hat{A}\hat{B}}\,\widehat{\Omega}_{\hat{A}}\,\widehat{\Omega}_{\hat{B}}\,\Big)\,.
 \ee 
We note that the $\widehat{\Omega}$ are separately gauge invariant under $\xi^{\hat{M}}$ transformations, 
but invariance under local Lorentz transformations requires the precise coefficients given here.

\subsubsection*{Kaluza-Klein ansatz and symmetries}

We now perform the Kaluza-Klein rewriting of the DFT action (\ref{DFToncemore}). Thus, 
we gauge fix the local Lorentz group as in (\ref{gaugeFIX}) and split the coordinates 
according to (\ref{COORdinatesplit}). Moreover, we assume that the 
fields are independent of $\tilde{x}_{\mu}$, but we stress that the dependence 
on $y$ and $\tilde{y}$ is still completely generic, and we write these coordinates collectively 
as $Y^M=(\tilde{y}_m,y^m)$. Therefore we have to split the indices according to  
 \be\label{indexsplit}
  {}^{\hat{M}} \ = \ \big(\,{}_{\mu}\,\,,{}^{\mu}\,\,,{}^{M}\,\big)\;, \qquad
  {}_{\hat{A}} \ = \ \big(\,{}^{a}\,,\,{}_{a}\,,\,{}_{A}\,\big) \;. 
 \ee 
With these index conventions the $O(D,D)$ metric reads 
 \be
  \eta_{\hat{M}\hat{N}} \ = \  \begin{pmatrix}    0 & \delta^{\mu}{}_{\nu} & 0 \\[0.5ex]
  \delta_{\mu}{}^{\nu} & 0 & 0 \\[0.5ex]  0 & 0 & \eta_{MN} \end{pmatrix}\;, 
 \ee 
where $\eta_{MN}$ denotes the $O(d,d)$ invariant metric. According to 
(\ref{vielbeinconstr}) the `flattened' $O(D,D)$ metric then takes the form 
  \be
   \eta_{\hat{A}\hat{B}} 
   \ = \ E_{\hat{A}}{}^{\hat{M}}E_{\hat{B}}{}^{\hat{N}}\eta_{\hat{M}\hat{N}} \ \equiv \ 
   \begin{pmatrix}    0 & \delta^{a}{}_{b} & 0 \\[0.5ex]
  \delta_{a}{}^{b} & 0 & 0 \\[0.5ex]  0 & 0 & \eta_{AB} \end{pmatrix}\;. 
 \ee 
In contrast, the $\delta$ symbol used in (\ref{genHdef}) is, by slight abuse 
of the usual terminology, defined as
  \be\label{deltaDef}
   \delta_{\hat{A}\hat{B}} \ = \ \begin{pmatrix}    \eta^{ab} & 0 & 0 \\[0.5ex]
  0 & \eta_{ab} & 0 \\[0.5ex]  0 & 0 & \delta_{AB} \end{pmatrix}\;, 
 \ee 
where $\eta_{ab}$ is the Lorentz metric of the $n$-dimensional, `external' tangent space 
and $\eta^{ab}$ its inverse.  This is the $O(D-1,1)\times O(D-1,1)$ invariant metric. 
We use the convention that curved indices $\hat{M},\hat{N}$ are raised and lowered 
with $\eta_{\hat{M}\hat{N}}$ and that flat indices $\hat{A}, \hat{B}$ are raised and lowered 
with $\eta_{\hat{A} \hat{B}}$. Let us stress that when splitting the indices in this formalism as in (\ref{indexsplit})
we have to carefully distinguish between upper and lower indices; they are not meant to be 
raised and lowered with the usual Lorentz metric $\eta_{ab}$, unless stated explicitly. 
Finally, we give the frame field in the Lorentz gauge fixed form appropriate for Kaluza-Klein 
compactifications, 
 \be\label{frameansatz}
  E_{\hat{A}}{}^{\hat{M}} \ = \ \begin{pmatrix}    E^{a}{}_{\mu} & E^{a\mu} & E^{aM} \\[0.5ex]
  E_{a \mu} & E_{a}{}^{\mu} & E_{a}{}^{M} \\[0.5ex]  E_{A\mu} & E_{A}{}^{\mu} & E_{A}{}^{M} \end{pmatrix}
  \ = \ \begin{pmatrix}    e_{\mu}{}^{a} & 0 & 0 \\[0.5ex]
  -e_{a}{}^{\nu} C_{\mu\nu} & e_{a}{}^{\mu} & -e_{a}{}^{\mu}A_{\mu}{}^{M} \\[0.5ex]  
  {\cal V}_A{}^{M} A_{\mu M} & 0 & {\cal V}_{A}{}^{M} \end{pmatrix}\;, 
 \ee 
where ${\cal V}$ is the vielbein for ${\cal H}_{MN}=({\cal V}{\cal V}^T)_{MN}$ and 
 \be\label{DEFC}
  C_{\mu\nu} \ = \ B_{\mu\nu}+\frac{1}{2}A_{\mu}{}^{M} A_{\nu M}\;.
 \ee
Here we have used the local Lorentz symmetry to set some components to zero. 
In this form the frame field satisfies the constraint (\ref{vielbeinconstr}) so that 
the inverse vielbein $E_{\hat{M}}{}^{\hat{A}}$ can be immediately written by raising and lowering indices. 
This form of the frame field is closely related to the one relevant for the 
heterotic theory \cite{Hohm:2011ex}, as it encodes additional gauge fields $A_{\mu}{}^{M}$, 
and also appears in an analogous form in Scherk-Schwarz reductions \cite{Geissbuhler:2011mx}.  
In addition, we redefine the dilaton according to 
 \be\label{dilaREDEF}
  e^{-2d} \ = \ e\, e^{-2\phi}\;,
 \ee
where $e$ denotes the determinant of the external vielbein $e_{\mu}{}^{a}$ 
defined in (\ref{frameansatz}). As we will see, this definition is such that $\phi$
is a scalar with respect to the external space but a density with respect to the 
internal space.     

Let us now turn to the discussion of the symmetries (\ref{fuilGAUGE}) 
acting on these Kaluza-Klein-type fields,   using  
the index split (\ref{indexsplit}) also for the gauge parameter,  
 \be
  \xi^{\hat{M}} \ = \ \big(- \Lambda_{\mu}\,,\,\xi^{\mu}\,,\,\Lambda^M\,)\;,  
 \ee
where we inserted a sign in order to comply with our conventions above.  
We first note that a compensating local Lorentz transformation 
with off-diagonal parameter $\lambda^{aB}$ 
is required in order to maintain the gauge choice in (\ref{frameansatz}). 
Using the condition $\tilde{\partial}^{\mu}=0$ we infer that the 
component $E^{a\mu}=0$ is left invariant under $\xi^{\hat{M}}$
transformations, but $E^{aM}$ and $E_{A}{}^{\mu}$ require 
the compensating local Lorentz transformation with parameter 
  \be\label{compparam}
   \lambda^{aA} \ = \  - {\cal V}_{M}{}^{A} \, \partial^M\xi^{\mu}\, e_{\mu}{}^{a}\;, \qquad
   \lambda_{A}{}^{a} \ = \ {\cal V}_{A}{}^{N}\partial_N\xi^{\nu} e_{\nu}{}^{a}\;.  
  \ee  
Since $\lambda_{\hat{A}}{}^{\hat{B}}$ should parametrize 
$O(D-1,1)\times O(D-1,1)$, we need to add  
additional parameters in order to preserve the invariant metric (\ref{deltaDef}).
We need to satisfy
 \be
  \lambda_{\hat{A}}{}^{\hat{C}}\,\delta_{\hat{C}\hat{B}}+ \lambda_{\hat{B}}{}^{\hat{C}}\,\delta_{\hat{A}\hat{C}} \ = \ 0\;, 
 \ee
which requires 
 \be
  \lambda_{a}{}^{A} \ = \ - \delta^{AB}{\cal V}_{B}{}^N\partial_N\xi^{\nu} e_{\nu a}\;, \qquad
  \lambda_{Aa} \ = \ \delta_{AB}{\cal V}_{M}{}^{B}\partial^M\xi^{\nu} e_{\nu a}\;.
 \ee

 Next, we apply the gauge transformations to components of the 
 frame field, starting with $E_{A}{}^{M}$, which yields
  \be
   \begin{split}
   \delta\,E_{A}{}^{M} \ = \ \delta\,{\cal V}_{A}{}^{M} \ &= \ \xi^{\hat{N}}\partial_{\hat{N}}{\cal V}_{A}{}^{M}
   +\big(\partial^M\xi_{\hat{N}}-\partial_{\hat{N}}\Lambda^M\big)E_{A}{}^{\hat{N}} +\lambda_{A}{}^{b}E_{b}{}^{M}\\
   \ &= \ \widehat{\cal L}_{\Lambda}{\cal V}_{A}{}^{M}+\xi^{\nu}\partial_{\nu}{\cal V}_{A}{}^{M}+\partial^M\xi^{\mu}\,
   {\cal V}_{A}{}^{N} A_{\mu N}-{\cal V}_{A}{}^{N}\partial_N\xi^{\mu} A_{\mu}{}^{M} \\
   \ &= \ \widehat{\cal L}_{\Lambda}{\cal V}_{A}{}^{M}+\xi^{\nu}D_{\nu}{\cal V}_{A}{}^{M}+\widehat{\cal L}_{\Lambda=\xi^{\nu}A_{\nu}}{\cal V}_{A}{}^{M}  \;.
  \end{split}
  \label{calcE}
  \ee  
 Here we have introduced the $\Lambda$-covariant derivative. 
Application to the component $E^{a}{}_{\mu}=e_{\mu}{}^{a}$ gives
after a completely analogous computation, 
 \be\label{deltae}
 \begin{split}
  \delta\,e_{\mu}{}^{a}  \ &= \ \Lambda^N\partial_N e_{\mu}{}^{a}+\xi^{\nu}\partial_{\nu}e_{\mu}{}^{a}
  +\big(\partial_{\mu}\xi^{\nu}-A_{\mu}{}^{N}\partial_N\xi^{\nu}\big)e_{\nu}{}^{a} \\
  \ &= \ \big(\Lambda^N+\xi^{\nu}A_{\nu}{}^{N}\big)\partial_Ne_{\mu}{}^{a}+\xi^{\nu}D_{\nu}e_{\mu}{}^{a}
  +D_{\mu}\xi^{\nu} e_{\nu}{}^{a}\;. 
 \end{split} 
 \ee  
This implies that $e_{\mu}{}^{a}$ is a $\Lambda$-scalar and we see that the transformation rule 
takes again the form of a 
(covariantized) diffeomorphism plus a $\Lambda$ transformation and a field-dependent 
transformation. 
In order to compute the gauge variation of $A_{\mu}{}^{M}$ we may evaluate the 
component $\delta_{\xi}E_{a}{}^{M}$, which equals  
  \be
   \delta\, E_{a}{}^{M} \ = \ -(\delta e_{a}{}^{\mu})A_{\mu}{}^{M}-e_{a}{}^{\mu}\delta A_{\mu}{}^{M}\;.
  \ee
 Using (\ref{deltae}) one finds for the $\Lambda^M$ variation, after some manipulations,  
  \be
   \delta_{\Lambda} A_{\mu}{}^{M} \ = \ \partial_{\mu}\Lambda^M+\big[\Lambda,A_{\mu}\big]_{D}^M+\partial^M\Lambda_{\mu}\;. 
  \ee
 We infer that this coincides precisely with the required gauge variation (\ref{betterdelta}), (\ref{oneformgauge}) of the 
 tensor hierarchy, including the shift-like symmetry parametrized by $\Lambda_{\mu}$,  
 which we now recover from the higher-dimensional diffeomorphisms of DFT. 
  Similarly, after some algebra the $\xi^{\mu}$ variation is found to be 
 \be 
  \delta_{\xi}A_{\mu}{}^{M} \ = \ \xi^{\nu}\partial_{\nu}A_{\mu}{}^{M}+\partial_{\mu}\xi^{\nu} A_{\nu}{}^{M}
  -A_{\mu}{}^{N}\partial_N\xi^{\nu} A_{\nu}{}^{M} + \partial^M\xi^{\nu} \,C_{\nu\mu}+{\cal H}^{MN}g_{\mu\nu} \partial_N\xi^{\nu}\;, 
 \ee  
where we used ${\cal H}^{MN}={\cal V}_{A}{}^{M}{\cal V}^{AN}$.  
 We can again rewrite this in a more covariant way by expressing it in terms of the 
 field strength and a field-dependent gauge transformation,  
   \be
  \begin{split}
   \delta_{\xi}A_{\mu}{}^{M} \ = \ &\, \xi^{\nu}{\cal F}_{\nu\mu}{}^{M}+{\cal H}^{MN}g_{\mu\nu}\partial_{N}\xi^{\nu} \\
   &+\delta_{(\Lambda=\xi^{\nu}A_{\nu})}A_{\mu}{}^{M}+\partial^M\big(-\xi^{\nu} C_{\mu\nu}\big)\;.
  \end{split}
  \label{calcA}
 \ee      
Comparing the transformation laws (\ref{calcE})--(\ref{calcA}) to the formulas derived in the previous section
we find complete agreement for the $\Lambda$ transformations. Moreover, 
we find for the diffeomorphisms of ${\cal V}$, $e$ and $A$ parametrized by $\xi^{\mu}$, 
 \be\label{covdelA}
 \begin{split} 
   \delta_{\xi}{\cal V}_{A}{}^{M} \ &= \  \xi^{\nu}D_{\nu}{\cal V}_{A}{}^{M}\;, \\
   \delta_{\xi} e_{\mu}{}^{a} \ &= \ \xi^{\nu}D_{\nu}e_{\mu}{}^{a}
  +D_{\mu}\xi^{\nu} e_{\nu}{}^{a}\;, \\
   \delta_{\xi}A_{\mu}{}^{M} \ &= \  \xi^{\nu}{\cal F}_{\nu\mu}{}^{M}+{\cal H}^{MN}g_{\mu\nu}\partial_{N}\xi^{\nu}\;, 
 \end{split} 
 \ee
up to a common field-dependent $\Lambda$ gauge transformation that can be ignored.  
Acting on the component $E_{a\mu}=-e_{a}{}^{\nu}C_{\mu\nu}$ reproduces by 
use of our above results (\ref{BTransF}) and $\Delta B_{\mu\nu}=2D_{[\mu}\Lambda_{\nu]}$,
as required, while we can now also determine the diffeomorphism variation parametrized by $\xi^{\mu}$. 
It is convenient to add right away gauge transformations with parameter 
$\Lambda^M=-\xi^{\nu}A_{\nu}{}^{M}$ and $\Lambda_{\mu}=\xi^{\nu}C_{\mu\nu}$ 
in order to obtain covariant diffeomorphisms. First one finds
 \be
  \delta_{\xi}C_{\mu\nu} \ = \ \xi^{\rho}\big(D_{\rho}C_{\mu\nu}+\partial_{\mu}C_{\nu\rho}-D_{\nu}C_{\mu\rho}-A_{\nu}{}^{N}\partial_{\mu}A_{\rho N}\big)
  +{\cal H}^{MN}\partial_N\xi^{\rho} g_{\rho\nu} A_{\mu M}\;,
 \ee 
which is not covariant. Next, we compute the covariant variation of $B_{\mu\nu}$, using 
 \be
  \Delta B_{\mu\nu} \ = \ \delta C_{\mu\nu}-A_{\mu}{}^{M}\,\delta A_{\nu M}\;, 
 \ee
which follows from (\ref{DeltaB}) and (\ref{DEFC}). After some algebra and employing (\ref{covdelA}) 
one finds the covariant form 
 \be
  \Delta_{\xi}\,B_{\mu\nu} \ = \   \xi^{\rho}{\cal H}_{\mu\nu\rho}\;. 
 \ee
Finally, applying the dilaton variation (\ref{dilvar}) to (\ref{dilaREDEF}) and using (\ref{deltae}) yields 
 \be
  \delta \, e^{-2\phi} \ = \ \xi^{\mu}D_{\mu}e^{-2\phi}+\partial_N\big(\Lambda^Ne^{-2\phi}\big)\;, 
 \ee 
up to the same field-dependent $\Lambda$ gauge transformation.  
Therefore, the dilaton is a scalar from the perspective of the external space, but a density from the internal space.

\subsubsection*{Kaluza-Klein action}
Let us now evaluate the DFT action for the Kaluza-Klein-type ansatz (\ref{frameansatz}).
To this end we compute the various components of the generalized 
coefficients of anholonomy (\ref{anholonomy1}) and (\ref{anholonomy2}). 
After some algebra one finds that various components vanish, 
 \be
  \widehat{\Omega}^{abc} \ = \ \widehat{\Omega}^{ab}{}_{c} \ = \ \widehat{\Omega}^{ab}{}_{C} 
   \ = \ \widehat{\Omega}^{a}{}_{BC} \ = \ \widehat{\Omega}^{a} \ = \ 0\;, 
 \ee 
while the non-vanishing components read
 \be\label{OMegacomp}
  \begin{split}
   \widehat{\Omega}^{a}{}_{bC} \ &= \ - {\cal V}_{C}{}^{M}e_{\mu}{}^{a}\partial_M e_{b}{}^{\mu}\;, \\ 
   \widehat{\Omega}_{ab}{}^{c} \ &= \ \Omega_{ab}{}^{c}\;, \\
   \widehat{\Omega}_{abc} \ &= \ e_{a}{}^{\mu}e_{b}{}^{\nu}e_{c}{}^{\rho}\, {\cal H}_{\mu\nu\rho}\;, \\
   \widehat{\Omega}_{abC} \ &= \ -e_{a}{}^{\mu} e_{b}{}^{\nu} \,{\cal F}_{\mu\nu}{}^{M}\,{\cal V}_{MC}\;,  \\
   \widehat{\Omega}_{aBC} \ &= \ e_{a}{}^{\mu} D_{\mu}{\cal V}_{B}{}^{M}\, {\cal V}_{MC}\;, \\
   \widehat{\Omega}_{ABC} \ &= \ \Omega_{ABC}\;, \\
   \widehat{\Omega}_{a} \ &= \ 
   \Omega_{ac}{}^{c}+2e_{a}{}^{\mu}D_{\mu}\phi\;, \\
   \widehat{\Omega}_{A} \ &= \ \Omega_{A}-{\cal V}_{A}{}^{M}\,e^{-1}\partial_M e\;.  
  \end{split}
 \ee  
Here we used the usual but $A$-covariantized coefficients of anholonomy, 
 \be
   \Omega_{ab}{}^{c} \ \equiv \ -2e_{[a}{}^{\mu} e_{b]}{}^{\nu}D_{\mu}e_{\nu}{}^{c} \quad \Rightarrow \quad
   \Omega_{ac}{}^{c} \ = \  -e^{-1}D_{\mu}\big(e\,e_{a}{}^{\mu}\big)\;, 
 \ee
and the generalized coefficients of anholonomy for the internal space, 
 \be
 \begin{split}
   \Omega_{ABC} \ &\equiv \ {\cal V}_{A}{}^{M}\partial_M{\cal V}_{B}{}^{N}\,{\cal V}_{NC}\;, \\
    \Omega_{A} \ &\equiv \ -e^{2\phi}\partial_M\big({\cal V}_{A}{}^{M}e^{-2\phi}\big)\;. 
 \end{split}
 \ee    
We observe that everything organizes immediately into the 
gauge covariant objects introduced in sec.~2.    
We can now evaluate the action by inserting (\ref{OMegacomp}) into (\ref{DFTLagrangian}). 
Again, after  some algebra one finds 
 \be
 \begin{split}
   {\cal L}_{\rm DFT} \ = \ e\,e^{-2\phi}\Big[\Big(&-\frac{1}{4}\Omega^{abc}\Omega_{abc}-\frac{1}{2}\Omega^{acd}\Omega_{adc}+\Omega_{ab}{}^{b}\,\Omega^{a}{}_{c}{}^{c}
   +4e^{a\mu}D_{\mu}\phi\,\Omega_{ac}{}^{c}\\
   &+e^{a\mu}e^{b\nu}{\cal F}_{\mu\nu}{}^{M} e_{a}{}^{\rho}\partial_M e_{\rho b}\Big)+4g^{\mu\nu}D_{\mu}\phi\,D_{\nu}\phi\\
   &+\frac{1}{4}g^{\mu\nu} D_{\mu}{\cal V}_{M}{}^{A}\,D_{\nu}{\cal V}_{A}{}^{M}
   -\frac{1}{4}g^{\mu\nu}{\cal H}_{MN}\delta^{AB}D_{\mu}{\cal V}_{A}{}^{M}\,D_{\nu}{\cal V}_{B}{}^{N}\\
   &-\frac{1}{12}{\cal H}^{\mu\nu\rho}{\cal H}_{\mu\nu\rho}-\frac{1}{4}{\cal H}_{MN} {\cal F}^{\mu\nu M}{\cal F}_{\mu\nu}{}^{N} \\ 
   &+\frac{1}{4}\delta^{AB}\Omega_{A}{}^{CD}\Omega_{BCD}-\frac{1}{12}\delta^{AB}\delta^{CD}\delta^{EF}\Omega_{ACE}\Omega_{BDF}
   +\delta^{AB}\Omega_{A}\Omega_{B}\\
   &-\frac{1}{2}g^{\mu\nu}{\cal H}^{MN}\eta_{ab}\partial_M e_{\mu}{}^{a}\,\partial_Ne_{\nu}{}^{b}
   +\frac{1}{2}{\cal H}^{MN} \partial_Me_{\mu}{}^{a}\,\partial_N e_{a}{}^{\mu}\Big]\;. 
 \end{split}
 \ee  
Let us now simplify this action and organize the terms in a more geometric fashion. First, 
we can combine the terms in the third line as 
 \be
  \frac{1}{8}g^{\mu\nu}D_{\mu}{\cal H}^{MN}D_{\nu}{\cal H}_{MN} \ = \   \frac{1}{4}g^{\mu\nu} D_{\mu}{\cal V}_{M}{}^{A}\,D_{\nu}{\cal V}_{A}{}^{M}
   -\frac{1}{4}g^{\mu\nu}{\cal H}_{MN}\delta^{AB}D_{\mu}{\cal V}_{A}{}^{M}\,D_{\nu}{\cal V}_{B}{}^{N}\;.
 \ee  
Note that the terms in the last line are exactly analogous, so we have 
 \be
   \frac{1}{4}{\cal H}^{MN}\partial_Mg^{\mu\nu}\,\partial_N g_{\mu\nu} \ = \    -\frac{1}{2}{\cal H}^{MN}\eta_{ab}\,g^{\mu\nu}\,\partial_M e_{\mu}{}^{a}\,\partial_Ne_{\nu}{}^{b}
   +\frac{1}{2}{\cal H}^{MN} \partial_Me_{\mu}{}^{a}\,\partial_N e_{a}{}^{\mu}\;.
 \ee  
The terms in the first line give, up to a total derivative, the covariantized Einstein-Hilbert term in string frame, 
 \be
      e^{-2\phi}eR \ = \ e^{-2\phi}e \left(-\frac{1}{4}\Omega^{abc}\Omega_{abc}-\frac{1}{2}\Omega^{acd}\Omega_{adc}+\Omega_{ab}{}^{b}\,\Omega^{a}{}_{c}{}^{c}
   +4e^{a\mu}D_{\mu}\phi\,\Omega_{ac}{}^{c}\right)+(\text{total derivative})\,, 
 \ee     
as one may verify by an explicit computation. 

Next, we note that  
the first term in the second line plays a role in order to 
maintain invariance of the 
Einstein-Hilbert term under local Lorentz transformations in presence of $\partial_M$ derivatives. 
Indeed, under the local Lorentz transformation  
 \be
  \delta_{\lambda}e_{\mu}{}^{a} \ = \ \lambda^a{}_{b} \, e_{\mu}{}^{b}\;, \qquad 
  \delta_{\lambda}\omega_{\mu}{}^{ab} \ = \ -D_{\mu}\lambda^{ab}\;, 
 \ee
we compute for the transformation of the Ricci scalar  
 \be\label{noninvR}
  \delta_{\lambda}R \ = \ \delta_{\lambda}\left(e_{a}{}^{\mu}e_{b}{}^{\nu}R_{\mu\nu}{}^{ab}\right)
  \ = \ - e_{a}{}^{\mu}e_{b}{}^{\nu} {\cal F}_{\mu\nu}{}^{M}\partial_M\lambda^{ab}\;, 
 \ee
where we used the non-trivial commutator (\ref{commF}) 
of $\Lambda$ covariant derivatives.  In order to repair this non-invariance we define the 
improved Riemann tensor
 \be
  \widehat{R}_{\mu\nu}{}^{ab} \ = \ R_{\mu\nu}{}^{ab}+ {\cal F}_{\mu\nu}{}^{M} e^{a\rho}\partial_M e_{\rho}{}^{b}\;, 
 \ee 
implying for the Ricci scalar
 \be
  \widehat{R} \ = \ R+e^{a\mu}e^{b\nu}{\cal F}_{\mu\nu}{}^{M} e_{a}{}^{\rho}\partial_M e_{\rho b}\;.
 \ee
This improved Ricci scalar is invariant under local Lorentz transformations 
thanks to the non-Lorentz invariance of $e^{-1}\partial_M e$, 
which cancels the variation in (\ref{noninvR}).

Finally, the total action can be written as  
   \be\label{finalAction}
  \begin{split}
   S \ = \ \int d^nx\, d^{2d}y\,e\,e^{-2\phi}\Big(  &\widehat{R}
   +4g^{\mu\nu}D_{\mu}\phi D_{\nu}\phi -\frac{1}{12}{\cal H}^{\mu\nu\rho}{\cal H}_{\mu\nu\rho}\\
   &+\frac{1}{8}g^{\mu\nu}D_{\mu}{\cal H}^{MN}D_{\nu}{\cal H}_{MN}
   -\frac{1}{4}{\cal H}_{MN}{\cal F}^{\mu\nu M}{\cal F}_{\mu\nu}{}^{N}-V\Big)\;,  
  \end{split}
 \ee  
where the potential is given by 
 \be\label{Potential}
  V(\phi,{\cal H},g) \ = \ -{\cal R}(\phi,{\cal H}) - \frac{1}{4}{\cal H}^{MN}\partial_Mg^{\mu\nu}\,\partial_N g_{\mu\nu} \;.
 \ee 
Note that since $g_{\mu\nu}$ is a $\Lambda$ scalar the extra term is separately $\Lambda$ gauge invariant.    
This coincides with the action given in the introduction.

\section{Remarks and outlook}   
In this paper we have presented a Kaluza-Klein-like form of double field theory, in which we decomposed 
fields and coordinates and gauge fixed the local Lorentz symmetry according to a $D=n+d$ split. 
In addition, we assumed that all fields are independent of the $n$ `non-compact winding coordinates'
$\tilde{x}_{\mu}$ but did not impose any truncation in the compact coordinates $Y^M=(\tilde{y}_m,y^m)$.
Consequently,  the action (\ref{finalAction}) is manifestly $O(d,d)$ invariant and  a convenient starting point for arbitrary 
Kaluza-Klein ans\"atze for both the conventional action (\ref{original}) and the DFT action. 
In particular, at no stage this construction requires a torus background/topology, but we may choose to 
specialize to this background.

Let us make a few remarks on the torus background.  
In this case we may expand the fields in Fourier modes, which reads, say, for the metric $g_{\mu\nu}$,
 \be
  g_{\mu\nu}(x,y,\tilde{y}) \ = \ \sum_{  m_n   =-\infty}^{\infty} \chi_{\mu\nu}^{ [m_n]}(x)\,e^{im_n y^n}
  +\sum_{w^n=-\infty}^{\infty} \psi_{\mu\nu}^{ [w^n]}(x)\,e^{iw^n \tilde{y}_n}\;. 
 \ee 
Here, the $\chi_{\mu\nu}$ are the Kaluza-Klein modes, with mode numbers $m_n$, and $\psi_{\mu\nu}$
 are the winding modes, with mode numbers $w^n$. As $g_{\mu\nu}$ is an $O(d,d)$ singlet, a T-duality 
 transformation simply acts as a transformation of the coordinate arguments $Y^M$. 
For instance, the T-duality exchanging $y$ and $\tilde{y}$ simply acts as
 \be\label{JODD}
  J \ \equiv \ \begin{pmatrix}    0 & {\bf 1} \\[0.5ex]
  {\bf 1} & 0 \end{pmatrix} \,\in \, O(d,d)\; :\qquad
  \chi_{\mu\nu}^{ [m_n]}  \quad \longleftrightarrow\quad \psi_{\mu\nu}^{ [w^n]}\;, 
 \ee 
thus exchanging Kaluza-Klein and winding modes. If we assume for simplicity 
that the internal $B$-field vanishes, so that the generalized metric ${\cal H}_{MN}$
depends only on the internal metric $G$, (\ref{JODD}) acts as 
 \be
  {\cal H}'(Y') \ = \ J\,{\cal H}(Y)\,J^T\quad \Rightarrow \quad G' \ = \ G^{-1}\;. 
 \ee 
Thus, in particular, if $G$ equals the background torus metric, T-duality inverts all
radii, exactly as one would expect. 
We should note that the Kaluza-Klein and winding modes are not independent. 
The weak constraint originating from the level-matching condition of string theory 
requires 
 \be
  \vec{m}\cdot \vec{w} \ = \ m_n\,w^n \ = \ 0\;.
 \ee 
 So far gauge invariance actually requires the strong constraint which implies 
 that only the Kaluza-Klein or the winding modes (or any combination related 
 via an $O(d,d)$ rotation) can be included. One may hope, however,  
 that the present formulation with its novel gauge structures and their close relation to 
 those in lower-dimensional gauged supergravity give hints of how to relax the 
 constraint.

The complete background independent 
theory discussed here can be viewed as some kind of `unbroken phase' 
for the theory of massive Kaluza-Klein or winding modes on a torus. 
The bare theory (without a specification of the background) 
has actually a continuous $O(d,d,\mathbb{R})$ symmetry, 
but once we specialize to a torus background we have non-trivial identifications, 
 $y \sim  y+2\pi ,\, \tilde{y}  \sim  \tilde{y}+2\pi$,   
which are only preserved by $O(d,d)$ transformations with integer-valued 
matrix entries. Thus, on a torus background the symmetry is broken to 
the discrete $O(d,d,\mathbb{Z})$, as is the case in the full string theory. 
Moreover, the flat torus background spontaneously breaks the gauge 
symmetry so that the higher modes of various fields become massive 
via some infinite-dimensional version of the Higgs mechanism. 
In fact, the potential induces a mass term for
the higher modes of $g_{\mu\nu}$, as can be seen by expanding 
 \be
  g_{\mu\nu}(x,Y) \ = \ \eta_{\mu\nu}+h_{\mu\nu}(x,Y)\;, 
 \ee
 and plugging this into the second term of (\ref{Potential}). Schematically, 
  \be
   V \ \sim \ {\cal H}^{MN}\partial_M h^{\mu\nu}\,\partial_N h_{\mu\nu}
    \ = \ G^{mn}\partial_m h^{\mu\nu}\,\partial_n h_{\mu\nu}+\cdots 
    \ \sim \ \sum_{n} |m_n|^2\, h^{(-n)\mu\nu}h_{\mu\nu}^{(n)}+\cdots\;, 
   \ee 
where we included only the Kaluza-Klein modes.\footnote{Here the mass term is not of the actual form required by Fierz-Pauli, 
but this is due to the string frame metric. After proper diagonalization of the 
kinetic terms, the correct Fierz-Pauli mass term emerges to lowest order.} 
Inclusion of the 
winding modes gives the analogous term, with mass terms involving  $|w^n|^2$ 
and contraction with the T-dual metric ${G}'_{mn}$. 
Similarly, the higher modes of $B_{\mu\nu}$ become massive due to the 
$B$-field modification of the field strength ${\cal F}_{\mu\nu}{}^{M}$, 
so that the (generalized) Yang-Mills term yields 
 \be
  -\frac{1}{4}{\cal H}_{MN}{\cal F}^{\mu\nu M}{\cal F}_{\mu\nu}{}^{N} \ = \ 
-\frac{1}{4}G^{mn}\partial_m B^{\mu\nu}\,\partial_n B_{\mu\nu}+\cdots  \ = \ 
  -\frac{1}{4}\sum_{n} |m_n|^2\, B^{(-n)\mu\nu}B_{\mu\nu}^{(n)}+\cdots \;, 
 \ee
inducing a mass term for the $B_{\mu\nu}^{(n)}$.   
More precisely, the higher modes of $g_{\mu\nu}$ and $B_{\mu\nu}$ will 
`eat' various components of (higher modes of) the Kaluza-Klein vectors $A_{\mu}{}^{M}$
and internal (scalar) components of ${\cal H}_{MN}$ in order for the counting of 
degrees of freedom to work out. 
We leave a more detailed evaluation of the form of the action for the massive 
modes and the Higgs mechanism to future work. 
Moreover, it would be very interesting to evaluate the gauge structures introduced 
here for the explicit mode expansion of fields and gauge parameters, introducing 
a higher-dimensional analogue of the Virasoro algebra, but now for a Courant-like 
algebraic structure rather than a Lie algebra. 

As stressed before, our action can be evaluated for an arbitrary Kaluza-Klein ansatz, 
in particular for a (generalized) Scherk-Schwarz ansatz. Reductions of this kind have 
already been discussed in the literature and it has been shown that the 
form of gauged supergravity written in the embedding tensor formalism   
naturally emerges \cite{Aldazabal:2011nj,Geissbuhler:2011mx,Grana:2012rr}. Consequently, the tensor hierarchy appears. The present paper
explains the emergence of these structure in terms of the higher-dimensional DFT 
geometry and gauge symmetries, without any actual reduction. 
We note that the generalized Scherk-Schwarz reductions 
cannot uplift all gauged supergravities to higher-dimensional DFT. 
Because of the strong constraint this is only possible for those theories that are 
T-dual to geometric reductions, while it is known that there are 
disjoint $O(d,d)$ orbits of consistent gaugings \cite{Dibitetto:2012rk}. Accordingly, there have been 
attempts to relax the strong constraint and thus to uplift all gauged supergravities \cite{Grana:2012rr,Geissbuhler:2013uka,Berman:2013uda}. 
While encouraging, it is fair to say that so far no entirely convincing proposal 
has appeared of how to reconcile such an ad-hoc relaxation with the 
gauge symmetries. In particular, to our knowledge no closed set of 
constraints replacing the strong constraint has been found that is consistent 
with the gauge symmetries. We believe that the present formulation may be well 
suited to address these problems, for instance along the lines of the similar 
case of massive type II theories \cite{Hohm:2011cp}. 
 
Another main motivation for the present paper has been the search for an approach that 
is further applicable to U-duality symmetries and theories such as 
11-dimensional supergravity. Although there have been a number of paper 
generalizing the DFT structures to various U-duality groups 
\cite{Berman:2010is,Berman:2011pe,Berman:2011cg,Berman:2011jh,Coimbra:2011ky,Berman:2012vc}, these are 
restricted to truncations of $D=11$ supergravity. Specifically, here one also performs a 
Kaluza-Klein-type decomposition of fields and coordinates but then truncates 
all external coordinates and off-diagonal field components and puts  
the external metric to the flat Minkowski metric. The reason for this rather 
severe truncation is that, in contrast to DFT, the truncated fields and coordinates 
do not naturally fit into an enlarged (higher-dimensional) generalized metric. 
Thus, for the moment the only way out seems to be a formulation that is not 
fully covariant in the sense that coordinates are split and treated on a 
different footing, but without imposing a truncation, as we have done in DFT in this paper. 
It turns out that in this way complete gravity theories can indeed be formulated in 
a manifestly U-duality covariant manner. In an accompanying paper 
we will show this for a $3+1$ decomposition of $D=4$ Einstein gravity, 
in which case the U-duality group is given 
by the Ehlers group $SL(2,\mathbb{R})$~\cite{Hohm:2013new2}. 
The emerging structures are closely related to those originating from DFT, 
suggesting various generalizations that should also be 
applicable to the complete 11-dimensional supergravity. 
Finally, it has recently been shown how to incorporate $\alpha'$ corrections 
into DFT \cite{Hohm:2013jaa}. 
If this approach is extendable to U-duality covariant formulations 
we would be able to compute higher-derivative M-theory corrections to 
11-dimensional supergravity in a U-duality covariant way.

\section*{Acknowledgments}
This work is supported by the
DFG Transregional Collaborative Research Centre TRR 33
and the DFG cluster of excellence "Origin and Structure of the Universe".
We would like to thank each others home institutions for hospitality.

%\bibliographystyle{JHEP2}
%\bibliography{refs}

\providecommand{\href}[2]{#2}\begingroup\raggedright\endgroup

   \end{document}